\documentclass[trackchanges,twocolumn]{aastex702}
\usepackage[normalem]{ulem} 
\usepackage{amsmath}
\usepackage{xcolor}

\begin{document}

\title{Probabilistic foreground diagnostics from dispersion measure excess in individual fast radio bursts}

\author[0009-0003-6465-8944]{Qi Zhou}
\affiliation{School of Physics and Astronomy, Beijing Normal University, Beijing 100875, China}
\email{202421101151@mail.bnu.edu.cn}  

\author[0009-0009-3255-4132]{Ran Gao}
\affiliation{Department of Astronomy, Tsinghua University, Beijing 100084, China}
\email[show]{gaor@mail.tsinghua.edu.cn}

\author[0000-0002-8492-4408]{Zhengxiang Li}
\affiliation{School of Physics and Astronomy, Beijing Normal University, Beijing 100875, China}
\affiliation{Institute for Frontiers in Astronomy and Astrophysics, Beijing Normal University, Beijing
102206, China}
\email{zxli918@bnu.edu.cn}

\author[0000-0002-3567-6743]{Zong-Hong Zhu}
\affiliation{School of Physics and Astronomy, Beijing Normal University, Beijing 100875, China}
\affiliation{Institute for Frontiers in Astronomy and Astrophysics, Beijing Normal University, Beijing
102206, China}
\affiliation{Department of Astronomy, School of Physics and Technology, Wuhan University, Wuhan 430072, China}
\email{zhuzh@bnu.edu.cn}

\begin{abstract}

Fast radio bursts (FRBs) provide a unique probe of diffuse baryons through their dispersion measure (DM), but interpreting a large DM for an individual burst is limited by the uncertain contribution from its host galaxy, which is hard to directly observed. We address this gap with a source-level diagnostic framework that treats $\mathrm{DM_{host}}$ as a latent variable drawn from a shared lognormal population, thereby marginalizing over the uncertain host contribution. Applying an expectation--maximization (EM) inference scheme to 92 localized FRBs, we jointly estimate the host-DM population and obtain a model-conditioned host-DM distribution for each burst. We propagate these distributions into an excess cosmological DM for each sightline, quantifying both its uncertainty and the probability that a positive excess remains after accounting for the host contribution. FRBs previously reported to show evidence for foreground or intervening contributions preferentially exhibit high probabilities of retaining such an excess, indicating that this source-level probability can provide a useful statistic for prioritizing foreground follow-up observations. Using this diagnostic, we identify 18 additional localized FRBs as high-priority candidates for follow-up searches for possible intervening galaxies, groups, clusters, or large-scale structure.

\end{abstract}


\section{Introduction} 

The spatial distribution of baryonic matter in the Universe represents one of the pivotal challenges in contemporary astrophysics and cosmology. In particular, the distribution of ionized gas within galaxy halos, the intergalactic medium (IGM), and the cosmic web constitutes a critical research frontier, as it is not only closely tied to galaxy formation and feedback processes but also fundamentally shapes our comprehension of large-scale structure formation and precision cosmological observations \citep{2017ARA&A..55..389T,Sorini_2022,connor_gas-rich_2025}. Fast radio bursts \citep{doi:10.1126/science.1147532} have emerged as a powerful diagnostic tool for addressing the ``missing baryon" problem, since their pulsed signals experience frequency-dependent dispersion delays induced by free electrons along the propagation path \citep{fukugita_cosmic_1998,shull_baryon_2012,li_cosmology-independent_2019}. Dispersion measure (DM) is defined as $\mathrm{DM} = \int n_e \, \mathrm{d}l$, the integral of free electron density along the line of sight, directly reflecting the integrated electron column density. As emphasized in subsequent investigations, the cosmological utility of FRBs stems precisely from their relative insensitivity to gas temperature and metallicity, enabling a comparatively direct probe of the total column density of diffuse cosmic gas without the complications that limit many alternative observational tracers \citep{macquart_census_2020,law_distant_2020}.

With the rapid accumulation of localized FRB samples \citep{hallinan2019dsa2000radiosurvey,https://doi.org/10.5281/zenodo.3765414,frbcollaboration2025chimefrboutriggersdesignoverview}, the DM--$z$ relation has now been established with sufficient statistical significance to reveal the substantial contribution of cosmic gas to signal propagation. Previous studies have employed the average Macquart relation to characterize the cosmological DM budget along FRB sightlines, with subsequent efforts attempting to decompose the total baryon inventory into contributions from the intergalactic medium (IGM), halo gas, and other constituent components. However, current methods primarily rely on population-averaged relations, leaving unresolved why an individual FRB with a high observed DM lies above the mean relation, since the excess may reflect the host-galaxy environment, foreground structural enhancement, or other unmodeled contributions. Because the host-galaxy DM is hard to measure directly, it should be treated as a latent variable and jointly inferred from the data, a task that population-averaged methods cannot achieve \citep{macquart_census_2020,james_z_2021,khrykin_flimflam_2024}.

In recent years, various approaches have been proposed or implemented to make use of this information, including cross-correlation between FRB positions and galaxy positions \citep{2021ApJS..257...59C,rafiei-ravandi_chimefrb_2021}; DM--galaxy cross-correlation \citep{2019PhRvD.100j3532M}; DM--shear cross-correlation \citep{2023arXiv230909766R}; and stacking analyses using large samples of unlocalized FRBs \citep{connor_observed_2022,2023ApJ...945...87W}. Another technique, known as foreground mapping, relies on extensive spectroscopic follow-up campaigns to reconstruct the three-dimensional cosmic web along each line of sight \citep{lee_constraining_2022,2025ApJS..277...64H}. With the release of DESI Data Release 1 \citep{2026AJ....171..285D} and the advent of large future spectroscopic galaxy surveys, such methods are expected to become increasingly prevalent---especially as the sample of localized FRBs simultaneously grows by orders of magnitude.

The foreground galaxy field has been shown to contain information beyond the average DM--$z$ relation, and this information is expected to grow in importance as the localized-FRB sample expands. In particular, \cite{hussaini_correlation_2025} demonstrated that excess cosmological DM is positively correlated with foreground galaxy number density, and that stacked excess-DM signals can extend to Mpc scales, suggesting that a substantial fraction of the relevant signal is associated with diffuse intergalactic structure rather than being dominated solely by the one-halo term of intervening galaxies. However, identifying promising foreground sightlines from the DM of individual FRBs remains challenging, limiting the systematic expansion of the still-small sample with reported foreground evidence. Here, we develop a model-conditioned, source-level diagnostic for this purpose.

Previous studies have estimated
the host-galaxy DM of individual FRBs using
H$\alpha$ emission-line observations 
\citep{2017ApJ...834L...7T,2025A&A...696A..81B}. Because homogeneous observational estimates of the host-galaxy DM are not available for the full sample, we treat the host contributions as latent variables drawn from a common lognormal population. This hierarchical model allows us to infer the host-DM population parameters while accounting for each source's contribution uncertainty.
However, a Gaussian residual with a single redshift-independent dispersion is not a satisfactory representation of the stochastic electron column along cosmological sightlines: the cosmological-DM distribution can be asymmetric, and its fractional width is expected to vary with redshift \citep{McQuinn_2014,macquart_census_2020,james_zdm_2022}.We therefore adopt a centered skew-normal conditional residual likelihood with a redshift-dependent standard deviation motivated by the relative-width scaling of cosmological-DM models. Using the expectation--maximization (EM) algorithm \citep{dempster_maximum_1977}, we then estimate the host-DM population parameters and the per-source host-DM distribution simultaneously. Each inferred distribution is propagated into an excess cosmological DM for the corresponding sightline, from which we construct the $P$--$z$ and $P$--$\Delta \mathrm{DM}_{\rm cos}$ planes for follow-up prioritization.

This paper is organized as follows. In Section 2, we present the DM decomposition, the $\mathrm{DM_{host}}$ population model, the EM inference scheme and the definition of the per-source residual cosmological DM, $\Delta \mathrm{DM}_{\rm cos}$. In Section 3, we present the fitting results, including the inferred $\mathrm{DM}_{\rm host}$ population distribution, the $P$--$z$ and $P$--$\Delta \mathrm{DM}_{\rm cos}$ planes, and the resulting high-priority FRB sightlines for foreground follow-up. In Section 4, we summarize our results, discuss two illustrative special cases, and highlight the limitations of the current method as well as possible directions for future improvement.

\section{Method}

\subsection{Expectation--maximization algorithm}
Expectation--Maximization (EM) algorithm \citep{dempster_maximum_1977} is used for maximum likelihood estimation (MLE) in the case of incomplete data (including missing values or hidden variables). The standard computational framework of the EM algorithm consists of alternating E-steps (Expectation-step) and M-steps (Maximization step). The convergence of the algorithm ensures that the iterations at least approach a local maximum.

The core idea of the EM algorithm is to optimize the likelihood function containing incomplete data by treating hidden variables. Assuming we have an observed dataset $X$ and hidden (unobserved) data $Z$, the goal is to estimate the model parameters $\theta$ through the observed data $X$. Due to the presence of hidden variables, the likelihood function is not directly tractable for optimization with respect to $X$:

\begin{equation}
L(\theta) = p(X \mid \theta) = \int p(X, Z \mid \theta)\, dZ.
\label{eq1}
\end{equation}

The key to the EM algorithm lies in separating the estimation of hidden variables and parameter updating into two steps, with each step being relatively simple:

E-steps (Expectation-step): given the current parameter $\theta^{(t)}$, calculate the posterior probability distribution $p\left(Z \mid X, \theta^{(t)}\right)$, and use it to estimate the expected value of the hidden variable (or its related quantity).

M-steps (Maximization step): using the expected value calculated in E-step, maximize the log-likelihood function of complete data with hidden variables to update model parameters $\theta$.

\subsection{Decomposition of dispersion measure}
The observed dispersion measure of fast radio bursts can be decomposed into the sum of several components along the line of sight. Defining a ``cosmological DM" as $\mathrm{DM_{cos}}$ $\equiv$ $\mathrm{DM_{IGM}}$ $+$ $\mathrm{DM_{halo}}$ \citep{macquart_census_2020,connor_gas-rich_2025,hussaini_correlation_2025}, it can be written as:

\begin{equation}
\text{DM}_{\mathrm{obs}} = \text{DM}_{\text{MW}} + \text{DM}_{\text{cos}} + \frac{\text{DM}_{\text{host}}}{1+z_{s}},
\label{eq:DMobs2}
\end{equation}
here, $\mathrm{DM}_{\rm obs}$ is the measured DM; 
$\mathrm{DM}_{\rm MW}$ is the total Milky Way contribution, including the Galactic disk contribution estimated from NE2001 \citep{cordes_ne2001i_2003} and a fixed halo contribution of $30~{\rm pc~cm^{-3}}$ adopted in this work; 
$\mathrm{DM}_{\rm cos}$ represents the contribution of the cosmological path outside the host galaxy, including the intergalactic medium and intervening halo structures; 
$\mathrm{DM}_{\rm host}$ represents the contribution of the host galaxy in the source rest frame, and the factor $(1+z_s)^{-1}$ converts this rest-frame contribution to the observed-frame DM.

In this work, we calculate the mean cosmological DM contribution based on the FRB's redshift:

\begin{equation}
\langle \text{DM}_{\text{cos}} \rangle = f_d f_e K \Omega_{\text{b}} H_0 \int_{0}^{z_{s}} \frac{(1+z)\,\text{d}z}{\sqrt{\Omega_{\Lambda} + \Omega_{\text{m}}(1+z)^3}},
\label{eq:DMcos}
\end{equation}
where $f_d$ denotes the fraction of baryons residing in the diffuse ionized medium. The constant $K$ is defined as ${K=3c/8{\pi}G{m_p}}$, where $c$ is the speed of light, $G$ is the gravitational constant, and ${m_p}$ is the proton mass. The parameters $\Omega_b$, $\Omega_m$ and $\Omega_\Lambda$ represent the cosmic baryon density, the matter density parameter, and the dark energy density parameter, respectively. We use the Planck 2018 results \citep{2020A&A...641A...6P}. The quantity $f_e$, taken to be 0.88, corresponds to the number of electrons per baryon, while $H_0$ denotes the Hubble constant. The term $f_d$ accounts for contributions from both intervening halo gas and the intergalactic medium (IGM), but excludes those from the host galaxy and the Milky Way. In this work, we take $f_d$=0.93 following \cite{connor_gas-rich_2025}.

In order to identify the additional contribution of line of sight relative to the average cosmological relationship, we define the excess cosmological dispersion for each FRB as \citep{hussaini_correlation_2025}:

\begin{equation}
\begin{split}
\Delta \mathrm{DM}_{\mathrm{cos},i}
&= \mathrm{DM}_{\mathrm{obs},i} - \mathrm{DM}_{\mathrm{MW},i} \\
&\quad - \langle \mathrm{DM}_{\mathrm{cos}}(z_i)\rangle
- \frac{\mathrm{DM}_{\mathrm{host},i}^{\mathrm{model}}}{1+z_i},
\end{split}
\label{eq:exc}
\end{equation}
where $\mathrm{DM}_{\mathrm{host},i}^{\mathrm{model}}$ represents the host contribution assigned to the i-th FRB under a given host model. This quantity essentially reflects the residuals of a single source relative to the average cosmological DM--$z$ relationship. 

When $\Delta \mathrm{DM}_{\mathrm{cos},i}>0$, it means that after deducting the Galactic contribution, the average cosmological contribution and the modeled host term, the FRB still retains a positive residual DM. Such a positive residual may indicate enhanced electron columns along the line of sight, for example from enhanced IGM structures, intervening galaxy halos, galaxy clusters, or other contributions not captured by the average background relation. On the contrary, if the quantity is negative, it means that the $\mathrm{DM_{obs}}$ of the source is lower than the average cosmological expectation under the given host model.

Unlike analyses that focus only on population-level statistical correlations, $\Delta \mathrm{DM}_{\mathrm{cos}}$ is defined on a source-by-source basis and can therefore be used to prioritize individual sightlines with unusually large positive residuals relative to the baseline model.

\subsection{Expectation--maximization inference with a
redshift-dependent skewed residual likelihood}
\label{2.2}

One of the important sources of uncertainty in the FRB dispersion-measure decomposition arises from the host-galaxy contribution. $\mathrm{DM}_{\mathrm{host}}$ can not be observed directly,but also has significant dispersion and skewness between different sources. In order to avoid using a uniform fixed value for all FRBs, we model the host-galaxy rest-frame DM as a single lognormal distribution \citep{macquart_census_2020}:
\begin{equation}
 \mathrm{DM}_{\mathrm{host}}\sim\operatorname{LogNormal}(\mu_h,\sigma_h),
 \label{eq:host_lognormal}
\end{equation}
where $\mu_h$ and $\sigma_h$ denote the location and width parameters in logarithmic space, respectively.


Cosmological simulations indicate that the DM distribution at fixed redshift can be positively skewed, with its shape depending on the gas distribution around halos and with massive halos contributing to the high-DM tail \citep{2023MNRAS.520.3626W}. Motivated by these results, we adopt a centered skew-normal likelihood to model fluctuations about the mean cosmological DM–redshift relation. This provides a flexible phenomenological approximation to the residual distribution:
\begin{equation}
\begin{aligned}
 p_{\mathrm{SN}}(\Delta \mathrm{DM}_{\mathrm{cos},i}\mid z_i,\Theta)
 &=\frac{2}{\omega_i}\,\phi(u_i)\,\Phi(\alpha_{\mathrm{SN}}u_i), \\
 u_i&=\frac{\Delta \mathrm{DM}_{\mathrm{cos},i}-\xi_i}{\omega_i},
\end{aligned}
\label{eq:sz_pdf}
\end{equation}
where $\phi$ and $\Phi$ are the standard normal probability density
and cumulative distribution functions. The parameter
$\alpha_{\mathrm{SN}}$ controls the asymmetry of the conditional
residual density. Defining
$\delta_{\mathrm{SN}}=\alpha_{\mathrm{SN}}/
\sqrt{1+\alpha_{\mathrm{SN}}^2}$ ~\citep{Azzalini1985ACO}, we use
\begin{equation}
 \omega_i=\frac{\sigma_{\mathrm{int}}(z_i)}{\sqrt{1-2\delta_{\mathrm{SN}}^2/\pi}},
 \qquad
 \xi_i=-\omega_i\delta_{\mathrm{SN}}\sqrt{\frac{2}{\pi}},
 \label{eq:sz_centering}
\end{equation}
which guarantees
$\mathbb{E}(\Delta \mathrm{DM}_{\mathrm{cos},i}\mid z_i)=0$ and
$\operatorname{Var}(\Delta \mathrm{DM}_{\mathrm{cos},i}\mid z_i)=\sigma_{\mathrm{int}}^2(z_i)$.

We model the redshift dependence of this conditional residual
standard deviation as
\begin{equation}
 \sigma_{\mathrm{int}}^2(z_i)\
 =\sigma_{\mathrm{floor}}^2+
 A_{\mathrm{eff}}^2\frac{\langle \mathrm{DM}_{\cos}(z_i) \rangle^2}{z_i}
 \qquad (z_i>0),
 \label{eq:sz_scatter}
\end{equation}
with $\sigma_{\mathrm{int}}(0)=\sigma_{\mathrm{floor}}$.
The redshift-dependent term is motivated by the
$\sigma_{\mathrm{DM}}=F/\sqrt{z}$ relative-width prescription
used for the non-Gaussian cosmological-DM distribution by
\cite{macquart_census_2020,james_zdm_2022}.
However, the literature parameter $F/\sqrt{z}$ describes the width
of a specific cosmological-DM probability density and is not
mathematically the standard deviation of our complete conditional
residual. Therefore $A_{\mathrm{eff}}$ is a freely fitted,
phenomenological amplitude rather than a measurement of the physical
feedback parameter $F$. The term $\sigma_{\mathrm{floor}}$ absorbs
residual variations not assigned to the redshift-dependent component.
Our likelihood is thus an effective skewed residual model, not a
simulation-calibrated reconstruction of the full cosmic-DM density.

For the i-th FRB, the host contribution is a latent variable rather than a direct observation, so its likelihood needs to be marginalized over $\mathrm{DM}_{\mathrm{host}}$. Denoting the model parameter set by $\Theta$,the likelihood of a single source can be written as: 
\begin{equation}
\begin{aligned}
 L_i(\Theta) &= \int
 p(\mathrm{DM}_{\mathrm{host},i}\mid\mu_h,\sigma_h)
 p_{\mathrm{Sz}}\!\left(\Delta \mathrm{DM}_{\mathrm{cos},i}\mid z_i,\Theta\right) \\
 &\quad \times d\mathrm{DM}_{\mathrm{host}, i},
\end{aligned}
\label{eq:sz_marginal}
\end{equation}
where the model parameters are
$\Theta=(\mu_h,\sigma_h,\sigma_{\mathrm{floor}},
A_{\mathrm{eff}},\alpha_{\mathrm{SN}})$.

We emphasize that no Gaussian distribution is assumed for the observed dispersion measure itself. Instead, $\mathrm{DM_{\rm obs}}$ enters the likelihood through the DM decomposition in Equation~\ref{eq:DMobs2}, with the stochastic contributions modeled separately. The observed-data likelihood of the full sample is
\begin{equation}
 \mathcal{L}(\Theta)=\prod_{i=1}^{N_{\mathrm{FRB}}}L_i(\Theta).
 \label{eq:sz_full_likelihood}
\end{equation}
We maximize this likelihood using the expectation--maximization
(EM) algorithm. In the E-step of iteration $t$, the conditional
host-DM posterior is
\begin{equation}
 q_i^{(t)}(\mathrm{DM}_{\mathrm{host},i})=
 \frac{p(\mathrm{DM}_{\mathrm{host},i}\mid\mu_h^{(t)},\sigma_h^{(t)})
 p_{\mathrm{Sz}}(\Delta \mathrm{DM}_{\mathrm{cos},i}\mid z_i,\Theta^{(t)})}
 {L_i(\Theta^{(t)})}.
 \label{eq:sz_estep}
\end{equation}
In the M-step, the lognormal host-population parameters have
analytic updates,
\begin{equation}
 \begin{split}
 \mu_h^{(t+1)}&=\frac{1}{N_{\mathrm{FRB}}}
 \sum_i\mathbb{E}_{q_i^{(t)}}[\ln \mathrm{DM}_{\mathrm{host},i}],\\
 (\sigma_h^2)^{(t+1)}&=\frac{1}{N_{\mathrm{FRB}}}
 \sum_i\mathbb{E}_{q_i^{(t)}}
 \left[(\ln \mathrm{DM}_{\mathrm{host},i}-\mu_h^{(t+1)})^2\right].
 \end{split}
 \label{eq:sz_host_mstep}
\end{equation}
The remaining parameters
$(\sigma_{\mathrm{floor}},A_{\mathrm{eff}},\alpha_{\mathrm{SN}})$
are updated by numerically maximizing the expected complete-data
log-likelihood of the residual density,
\begin{equation}
 \sum_i\mathbb{E}_{q_i^{(t)}}
 \left[\ln p_{\mathrm{Sz}}\bigl(\Delta \mathrm{DM}_{\mathrm{cos},i}\mid z_i,\Theta\bigr)\right].
 \label{eq:sz_residual_mstep}
\end{equation}
We evaluate the host marginalization on a fixed grid in
$\ln \mathrm{DM}_{\mathrm{host},i}$ using numerical quadrature, and use multiple initial
values for the skewness parameter to reduce sensitivity to local
maxima. 
The marginalization over the latent host DM is performed
numerically using composite trapezoidal quadrature on a
uniformly spaced grid in $t=\ln \mathrm{DM}_{\mathrm{host}}$, with 1201 nodes spanning
$[-6,15]$. This transformation allows the lognormal host
density to be evaluated as a Gaussian density in log-space.
The same quadrature scheme is employed to evaluate the
observed-data likelihood and the source-specific conditional
distributions in the E-step.

In the M-step, the host-population parameters $\mu_h$ and
$\sigma_h$ admit closed-form updates in terms of the posterior
moments of $\ln \mathrm{DM}_{\mathrm{host}}$, whereas the residual-distribution parameters
$(\sigma_{\rm floor}, A_{\rm eff}, \alpha_{\rm SN})$ are updated
numerically using the L-BFGS-B algorithm. The marginal
likelihood is evaluated using log-sum-exp operations for
numerical stability. 

We also assess the sensitivity of the
likelihood and source-level probabilities to a finer integration
grid of 1801 nodes. We further assess the numerical convergence by
increasing the integration grid from 1201 to
1801 nodes over the same interval. The maximum
absolute difference in the source-level
excess-DM probabilities is only
$3.54\times10^{-5}$, while the negative
log-likelihood remains unchanged at the
reported numerical precision. These results
indicate that the adopted quadrature resolution
is sufficient for our analysis.

Iterations are stopped when the change in the observed-data
negative log-likelihood falls below the adopted convergence
tolerance. The resulting maximum-likelihood parameter estimates
are subsequently used to compute the source-specific host-DM
posterior distributions.

\subsection{Source-level excess-DM posteriors and probability ranking}
Since $\mathrm{DM}_{\mathrm{host},i}$ is a probabilistic inference rather than a deterministic value, $\Delta \mathrm{DM}_{\mathrm{cos},i}$ should also be treated as a posterior distribution rather than a single value. To this end, this paper propagates the posterior uncertainty of each FRB’s host DM into its corresponding excess cosmological DM.  

We characterize the median value and the asymmetric credible intervals of $\Delta \mathrm{DM}_{\mathrm{cos},i}$ using the 16th, 50th, and 84th percentiles of the posterior distribution:$\Delta \mathrm{DM}_{\mathrm{cos},i}^{\mathrm{p16}}$, $\Delta \mathrm{DM}_{\mathrm{cos},i}^{\mathrm{p50}}$, $\Delta \mathrm{DM}_{\mathrm{cos},i}^{\mathrm{p84}}$. Compared to symmetric Gaussian error bars, this quantile-based representation is more suitable for the problem at hand, as the prior and posterior of the $\mathrm{DM_{host}}$ are often significantly skewed, a skewness that naturally propagates into the posterior of $\Delta \mathrm{DM}_{\mathrm{cos}}$. Therefore, summarizing the uncertainty in terms of percentile intervals provides a more faithful reflection of the model outputs.

In addition to the quantile interval, this paper also calculates the posterior probability that each FRB meets the positive excess cosmological DM:

\begin{equation}
 \begin{split}
 P_i&\equiv P(\Delta\mathrm{DM}_{\mathrm{cos},i}>0
 \mid\mathrm{DM}_{\mathrm{obs},i},z_i,\Theta)\\
 &=\int_{\Delta DM_{\text{cos}} > 0}
 p(\mathrm{DM}_{\mathrm{host},i}\mid\mathrm{DM}_{\mathrm{obs},i},z_i,\Theta)\,d\mathrm{DM}_{\mathrm{host},i}.
 \end{split}
 \label{eq:sz_positive_probability}
\end{equation}

This quantity can be used as a compact statistical indicator of whether an FRB is a foreground-excess candidate under the a dopted model.

Moreover, the EM algorithm itself does not directly output the uncertainty of hyperparameters or model parameters (such as standard errors or confidence intervals). The EM algorithm is only responsible for providing point estimates, and uncertainty requires additional calculation afterwards. To quantify the statistical uncertainty related to the parameters in the population parameter ${\Theta}$, we used Markov Chain Monte Carlo (MCMC) analysis. The posterior constraints obtained are in Appendix~\ref{app:mcmc}.

\section{RESULTS}
\subsection{Fitting results}
\begin{figure}[htbp]
    \centering
    \includegraphics[width=0.95\columnwidth]{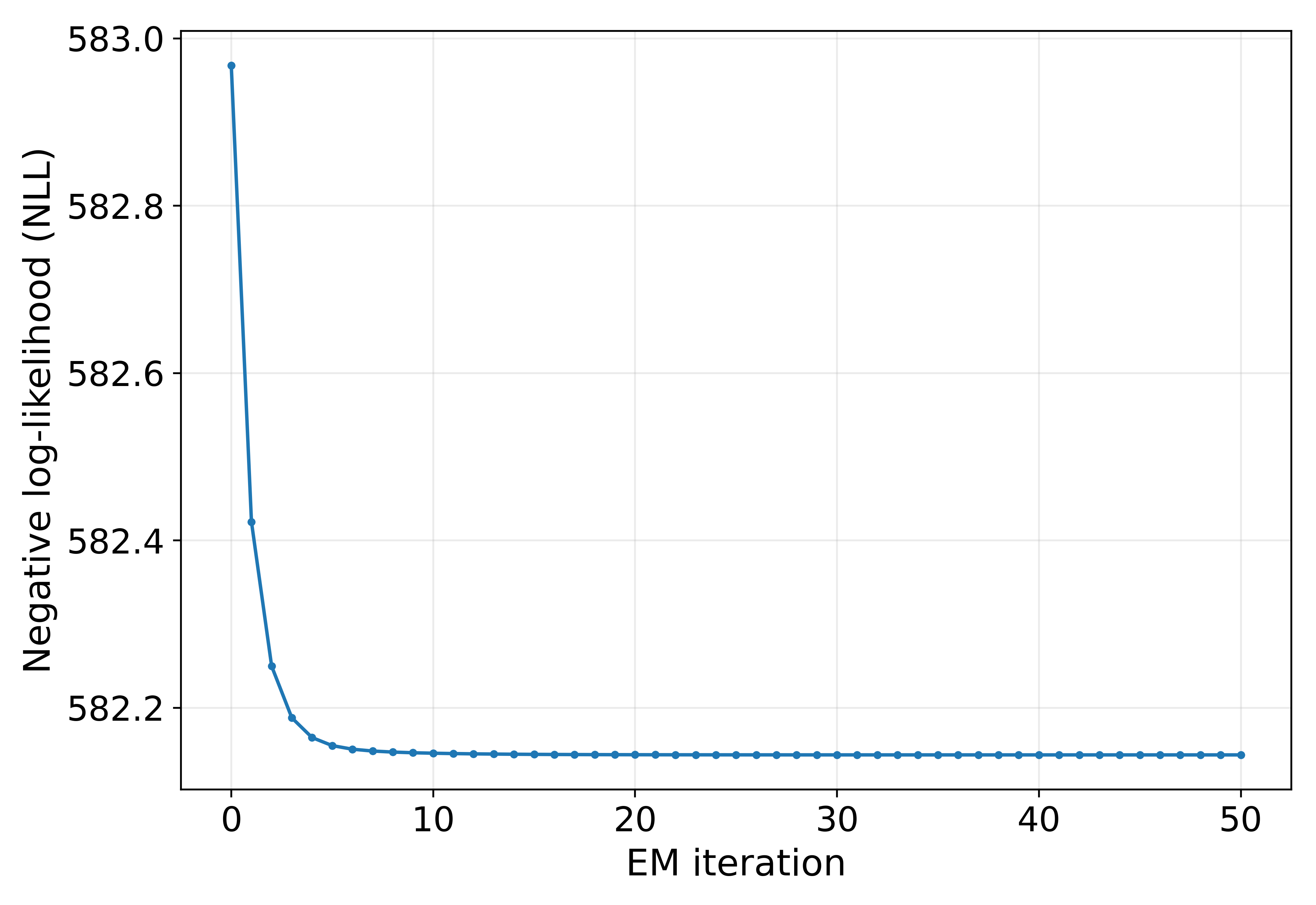}
    \caption{Convergence of the EM algorithm. The negative log-likelihood decreases with iteration number and stabilizes after 50 iterations.}
    \label{fig:em}
\end{figure}

Figure~\ref{fig:em} illustrates the convergence trajectory of the EM algorithm over successive iterations. The x-axis denotes the iteration number, while the y-axis represents the negative log-likelihood (NLL). As shown in the figure, the NLL value declines sharply during the initial iterations, followed by a gradual reduction, and eventually it decreases monotonically and
converges after 50 iterations to
$\mathrm{NLL}=582.144$. The fitted host-population parameters are
$\mu_h=4.5756$ and $\sigma_h=0.9642$. This pattern indicates that the EM algorithm converges to a local optimum of the observed-data likelihood.

\begin{figure}[htbp]
    \centering
    \includegraphics[width=0.95\columnwidth]{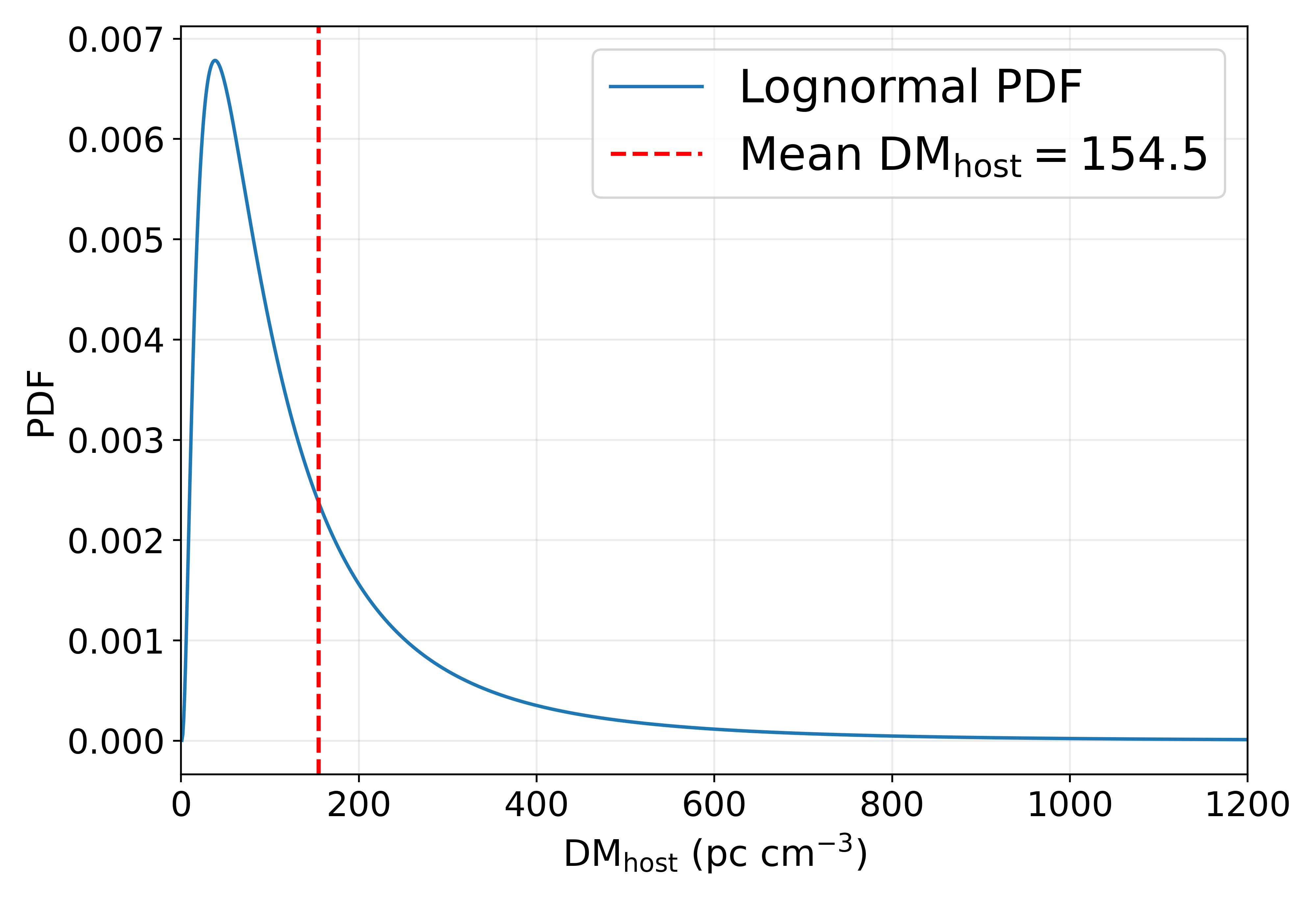}
    \caption{The host-DM probability density distribution fitted from the 92 localized FRBs. The blue solid line shows the single-lognormal PDF, and the the red dashed line marks the mean of the fitted lognormal population.}
    \label{fig:pdf2}
\end{figure}

Our analysis is based on 92 localized FRBs compiled from \cite{wang_probing_2025}. We directly fit these 92 localized FRBs, and the inferred host-DM population is shown in Figure~\ref{fig:pdf2}. The fitted mean host contribution is \(154.5~{\rm pc\ cm^{-3}}\), broadly consistent with the commonly adopted value \(\langle \mathrm{DM}_{\rm host}\rangle=150~{\rm pc\ cm^{-3}}\) used in previous studies \citep{james_z_2021,khrykin_flimflam_2024,connor_gas-rich_2025}. This fitted population provides the baseline host-DM model used to marginalize over the uncertain host contribution and to propagate the resulting inference into the per-source residual \(\Delta \mathrm{DM}_{\rm cos}\). The skewed lognormal shape further motivates using percentile intervals rather than a single fixed host-DM value. In addition, we obtain
$\sigma_{\mathrm{floor}}=51.75\,\mathrm{pc\,cm^{-3}}$,
$A_{\mathrm{eff}}=0.13589$, and
$\alpha_{\mathrm{SN}}=2.1125$. The fitted positive skewness
allows the residual likelihood to be asymmetric, while the
conditional residual standard deviation increases with redshift. To quantify the statistical uncertainties in the
fitted parameters, we additionally perform
an MCMC analysis using the same marginalized
likelihood as adopted in the EM framework.
The resulting posterior medians and corresponding
68\% credible intervals are
$\mu_h=4.570^{+0.186}_{-0.195}$,
$\sigma_h=0.987^{+0.176}_{-0.168}$,
$\sigma_{\rm floor}
=52.523^{+20.893}_{-17.254}\,
\mathrm{pc\,cm^{-3}}$,
$A_{\rm eff}=0.141^{+0.030}_{-0.029}$,
and
$\alpha_{\rm SN}=1.406^{+3.337}_{-2.831}$.
These posterior central values are broadly
consistent with the maximum-likelihood estimates
obtained from the EM algorithm.
However, the relatively broad posterior of
$\alpha_{\rm SN}$ indicates that the residual
skewness remains weakly constrained by the
current sample.
The full marginalized posterior distributions and parameter correlations are presented in Appendix~\ref{app:mcmc}.

\subsection{Model-conditioned ranking}
We use $P(\Delta \mathrm{DM}_{\rm cos}>0)$, the uncertainty interval of $\Delta \mathrm{DM}_{\rm cos}$, and the redshift of each FRB in the $P$--$z$ and $P$--$\Delta \mathrm{DM}_{\rm cos}$ planes to define a continuous follow-up priority. For comparison, we compile nine localized FRBs for which previous studies have reported evidence for possible foreground or intervening contributions:
FRB190520B \citep{Lee_2023}, FRB190614D \citep{law_distant_2020}, FRB190714A \citep{simha_searching_2023}, FRB200430A and FRB200906A \citep{simha_searching_2023}, FRB220914A \citep{sherman_deep_2024}, FRB221029A \citep{connor_gas-rich_2025}, FRB240114A \citep{chen_host_2025} and FRB240123A \citep{connor_gas-rich_2025}.

\begin{figure}[htbp]
    \centering
    \includegraphics[width=1.0\columnwidth]{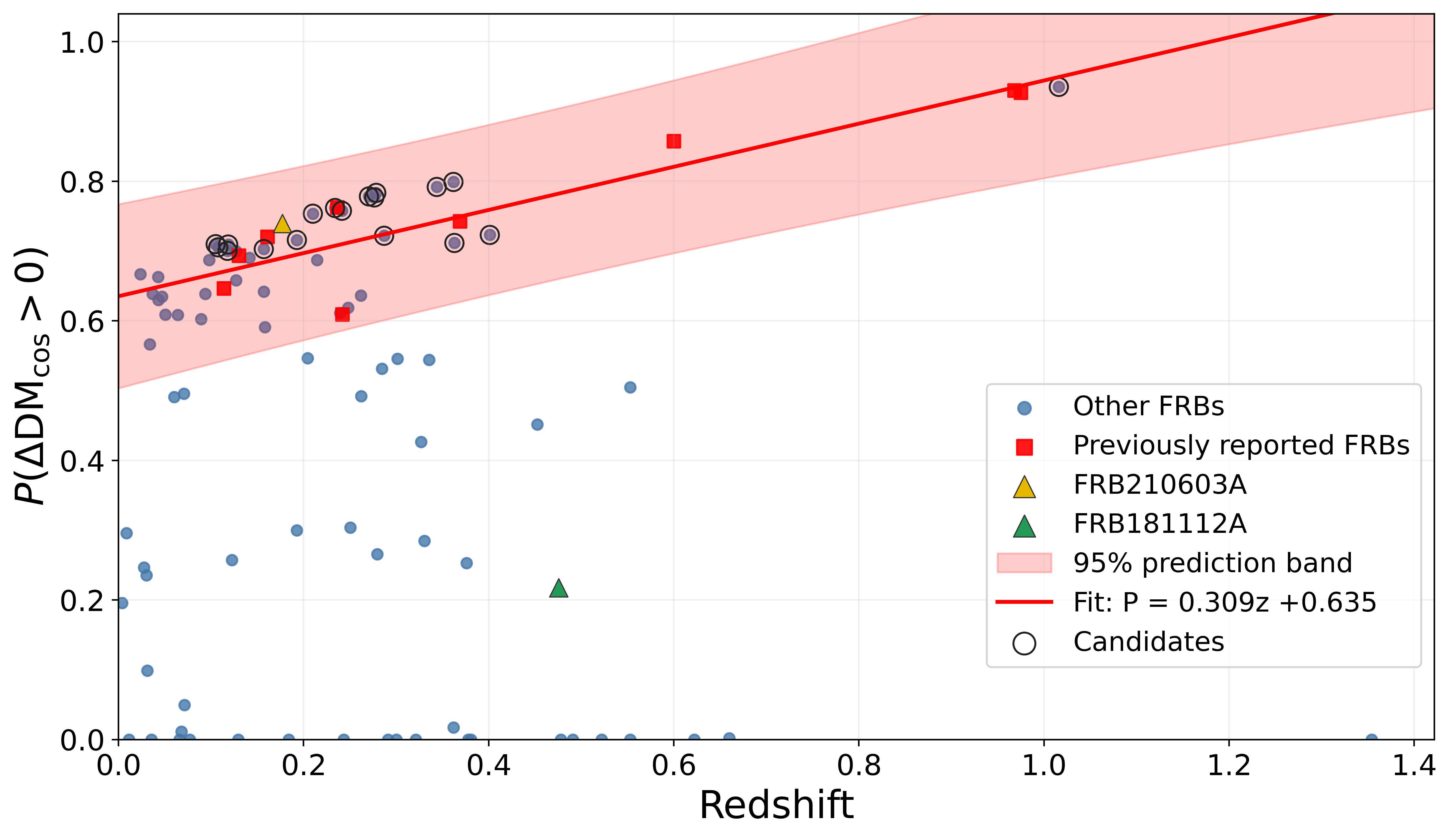}
    \caption{The model-conditioned probability $P(\Delta \mathrm{DM}_{\rm cos}>0)$ as a function of redshift. Red squares represent FRBs with reported evidence for possible foreground or intervening contributions; yellow and green triangles mark FRB210603A and FRB181112A, respectively; blue dots represent the remaining FRBs. The red solid line and the shaded region show the empirical $P$--$z$ trend fitted to the nine reported-foreground sources and its 95\% band; given the small reported-foreground comparison sample, we treat it as a visual guide rather than a formal prediction band.}
    \label{fig:pz}
\end{figure}

We present a continuous diagnostic view of the sample. Figure~\ref{fig:pz} shows the model-conditioned probability $P(\Delta \mathrm{DM}_{\rm cos}>0)$ as a function of redshift. The red squares in the figure represent FRBs with reported evidence for possible foreground or intervening contributions; yellow and green are two exceptions (see Section~\ref{sec:discussion} for details), and blue represent other FRBs. The red solid line and the shaded region show the empirical $P$--$z$ trend fitted to the nine reported-foreground sources and its 95\% band \citep{montgomery2021linear}, In addition, we also circle the candidates with black circles.

In the redshift space, the nine reported-foreground sources trace a roughly linear locus, providing the empirical fitting relationship,
\begin{equation}
    P=0.309z+0.635,
\end{equation}
together with a 95\% band. We emphasize that this relation is based on only nine points, whose $P$ values are themselves derived from the same host-DM model; we therefore treat it only as a qualitative visual guide for locating high-$P$ regions, not as a formal predictive relation.

\begin{figure}[htbp]
    \centering
    \includegraphics[width=1.0\columnwidth]{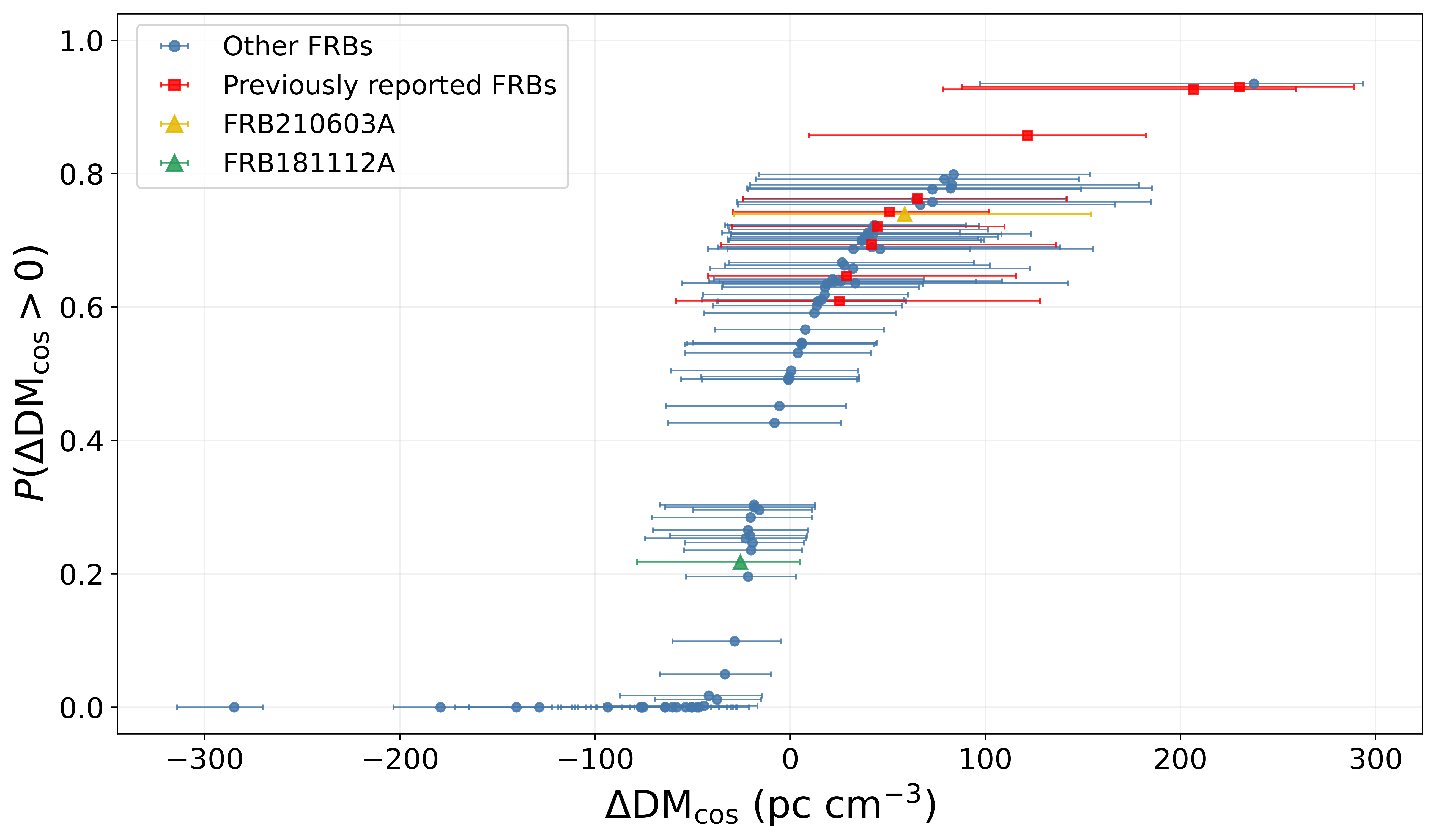}
    \caption{The model-conditioned probability $P(\Delta \mathrm{DM}_{\rm cos}>0)$ versus the inferred excess cosmological DM $\Delta \mathrm{DM}_{\mathrm{cos}}$, with per-source uncertainty intervals. Red squares represent FRBs with reported evidence for possible foreground or intervening contributions; yellow and green triangles mark FRB210603A and FRB181112A, respectively; blue dots represent the remaining FRBs. The gray horizontal line is the inferred interval from the 16th percentile to the 84th percentile, and the gray vertical line represents $\Delta\mathrm{DM}_{\mathrm{cos}}^{\mathrm{p50}}$.}
    \label{fig:pDM}
\end{figure}

Figure~\ref{fig:pDM} shows the model-conditioned probability that each FRB exhibits a positive excess cosmological DM together with its inferred uncertainty interval. The red squares represent FRBs with reported evidence for possible foreground or intervening contributions; yellow and green are two exceptions (see Section~\ref{sec:discussion} for details), and blue represent other FRBs. The gray horizontal line is the inferred interval from the 16th percentile to the 84th percentile, and the gray vertical line represents $\Delta\mathrm{DM}_{\mathrm{cos}}^{\mathrm{p50}}$.

Figure~\ref{fig:pDM} further compares $P(\Delta \mathrm{DM}_{\rm cos}>0)$ with the inferred distribution of $\Delta \mathrm{DM}_{\rm cos}$. This plane provides a more informative follow-up diagnostic than a discrete classification scheme, because it preserves both the sign and the uncertainty of the inferred excess DM. Sources with high $P$ and positive $\Delta \mathrm{DM}_{\rm cos}$ intervals constitute the most promising targets for foreground-structure searches, whereas sources with low $P$ or negative median $\Delta \mathrm{DM}_{\rm cos}$ provide useful control sightlines. In this sense, these planes translate the per-source excess-DM inference into a model-conditioned observational prioritization scheme. 

Guided by the locus of FRBs with reported foreground evidence, we define a follow-up candidate sample by selecting FRBs with $P > 0.7$. Because the inferred distribution is continuous, this threshold already implies a positive median ($\Delta \mathrm{DM}_{\mathrm{cos}}^{\mathrm{p50}} > 0$), so the median condition is not applied as an independent criterion. We deliberately do not rely on the empirical $P$--$z$ band for this selection, given the small reported-foreground comparison sample and the model-dependent origin of $P$. These sources are not claimed to have confirmed foreground or intervening contributions; rather, they represent systems that occupy the same high-probability region of the $P$--$z$ and $P$--$\Delta \mathrm{DM}_{\rm cos}$ planes as FRBs with reported foreground evidence. They are therefore suitable targets for follow-up observations aimed at testing possible line-of-sight or host-environment contributions, such as deeper imaging and spectroscopy of the host field, searches for intervening galaxies, and improved constraints on the local and foreground DM budget. In addition, their locations in the $P$--$\Delta \mathrm{DM}_{\rm cos}$ plane suggest that they occupy a parameter space similar to sources with reported foreground evidence, providing an additional motivation for follow-up.

Following this criterion, and excluding the sources already reported as showing foreground or intervening evidence as well as the two special cases discussed in Section~\ref{sec:discussion}, we prioritize 18 FRBs for further observation: FRB121102A, FRB190608B, FRB191001A, FRB191106C,
FRB211203C, FRB220105A, FRB220204A, FRB220610A,
FRB220717A, FRB220726A, FRB220825A, FRB221116A,
FRB230307A, FRB230526A, FRB230708A, FRB240213A,
FRB240215A, and FRB240229A.

The threshold is a practical ranking convention rather than a physical detection significance. For example, the probabilities of FRB190608B and FRB230628A are approximately 0.7003 and 0.6998, respectively, and hence lie on opposite sides of the threshold despite being nearly identical in probability. The candidate list should therefore be interpreted as a model-conditioned set of follow-up priorities, not as a confirmed sample of foreground halos or an exhaustive census of intervening structures.

\section{CONCLUSION AND DISCUSSION} 
\label{sec:discussion}

We extend population-level statistical analyses of localized FRBs to a source-by-source diagnostic framework. We first constrain the plausible range of $\mathrm{DM}_{\rm host}$ using the information from the full sample, and then propagate the inferred distribution of $\mathrm{DM}_{\rm host}$ into $\Delta \mathrm{DM}_{{\rm cos}}$ for each individual sightline. This residual is a model-conditioned quantity: it is defined relative to the adopted mean $\mathrm{DM}_{\rm cos}$--$z$ relation and the assumed host-DM model. Rather than uniquely distinguishing between host-galaxy and foreground contributions, our framework quantifies the model-conditioned probability that an individual FRB retains a positive DM residual after accounting for the inferred host contribution and the mean cosmological DM–redshift relation. This probability can serve as a statistical diagnostic for identifying promising sightlines for subsequent foreground investigations. The resulting probabilities and uncertainty intervals of $\Delta \mathrm{DM}_{\rm cos}$ provide a practical way to prioritize targets for subsequent multi-wavelength follow-up observations. In this sense, our framework provides a more refined source-by-source tool for using FRBs to study the cosmic web and the baryon distribution. Our results are consistent with the emerging picture that excess $\mathrm{DM}$ in FRBs arises from intervening large-scale structure along the line of sight in addition to host-galaxy contributions.

We emphasize that the target priorities provided by this method do not directly correspond to the magnitude of the foreground contribution. In the present analysis, several sources of uncertainty are not explicitly modeled, including the diversity of host-galaxy environments and systematic differences among Galactic electron-density models. These effects may distort $P(\Delta \mathrm{DM}_{\mathrm{cos}}>0)$, especially for low-redshift FRBs where the cosmological DM contribution is relatively small. Therefore, the proposed statistic should be interpreted as a follow-up prioritization tool rather than a definitive measure of the true foreground contribution.

We have additionally quantified the statistical uncertainties of the five model parameters through an MCMC analysis, as presented in Appendix~\ref{app:mcmc}. The resulting posterior constraints are broadly consistent with the EM maximum-likelihood estimates, while revealing non-negligible parameter correlations and a relatively weak constraint on the residual skewness parameter. This analysis provides a quantitative assessment of the statistical uncertainty associated with the fitted population model. Nevertheless, the remaining systematic uncertainties, including the adopted Galactic electron-density model, Milky Way halo contribution, and possible diversity of host-galaxy DM populations, are not included in the present MCMC analysis and require further investigation.

We further discuss two special cases that illustrate the limitations of adopting a single host-DM population model. For FRB210603A, the relatively large inferred $\mathrm{DM}_{\rm host}$ may be explained by a long path length through the disk of its host galaxy along the line of sight \citep{cassanelli_fast_2024}. This case highlights the potential impact of host-galaxy diversity on the inferred residuals; future work should consider host-dependent DM populations in order to obtain more accurate source-level probabilities. FRB181112A, in contrast, lies close in projection to a massive foreground galaxy, but this galaxy has been shown not to dominate the observed DM contribution \citep{prochaska_low_2019}. In our framework, such a sightline can be accommodated by the intrinsic-scatter term $\sigma_{\rm int}$ rather than being identified as a strong foreground-excess candidate. These two examples emphasize that $\Delta \mathrm{DM}_{\rm cos}$ should be interpreted as a model-conditioned residual, not as a direct measurement of the foreground DM contribution. Its interpretation remains subject to uncertainties in the $\mathrm{DM}_{\rm host}$ distribution, systematic differences among Galactic electron-density models, and stochastic fluctuations in the intergalactic medium. In particular, variations in halo gas and filamentary structures may introduce additional scatter that is not explicitly modeled in the present framework.

Based on the above discussion, although $\Delta \mathrm{DM}_{\rm cos}$ can be used as a model-conditioned diagnostic of possible foreground structure, a detailed decomposition of the relevant DM contributions and more accurate modeling of the IGM, halo gas, and host-galaxy environments remain important. Optical observations of host galaxies will also be essential for constraining host associations, redshifts, galaxy properties, and local environments, thereby helping to distinguish host-dominated high-DM events from sightlines affected by intervening structures. 

The assumed Milky Way halo contribution introduces an additional systematic uncertainty. We adopt a fixed value of 30 pc cm$^{-3}$ following \cite{connor_gas-rich_2025}, who explored halo contributions of 10 and 50 pc cm$^{-3}$ and found that the inferred cosmic-gas parameters changed only at approximately the percent level. That robustness was established at the population level, however, and is not directly transferable to the source-level probabilities presented here, which can be more sensitive to the adopted halo contribution, especially for low-redshift sightlines; this uncertainty is therefore not propagated in the present analysis. Future work combining this source-level diagnostic with optical host-galaxy observations, foreground galaxy surveys, and large-scale structure reconstructions will enable more direct identification of the baryonic content along individual FRB sightlines.

\begin{acknowledgments}
We are grateful to Professor He Gao for his helpful discussions. This work was supported by the National Key Research and Development Program of China Grant Nos. 2024YFC2207400, 2023YFC2206702, 2021YFC2203001; the National Natural Science Foundation of China under Grants Nos. 12322301, 12222302, 12275021 and 12405050; the China Postdoctoral Science Foundation under Grant Number 2026M793508; the Strategic Priority Research Program of the Chinese Academy of Sciences, Grant No. XDB23040100, the Interdiscipline Research Funds of Beijing Normal University, and the Fundamental Research Funds for the Central Universities.
\end{acknowledgments}

\appendix
\section{MCMC Constraints on Model parameters}
\label{app:mcmc}

To quantify the statistical uncertainties
associated with the parameters inferred
from the expectation--maximization (EM)
algorithm, we perform an additional Markov
chain Monte Carlo (MCMC) analysis using the
same redshift-dependent skew-normal likelihood
described in Section~\ref{2.2}.

The five model parameters are
\begin{equation}
{\Theta}
=
(\mu_h,\sigma_h,\sigma_{\rm floor},
A_{\rm eff},\alpha_{\rm SN}).
\end{equation}

The host-galaxy DM contributions are treated
as latent variables and marginalized over
numerically at each MCMC step.
The posterior distribution of the
parameters is therefore given by
\begin{equation}
p({\Theta}\mid\mathcal{D})
\propto
p({\Theta})
\prod_{i=1}^{N_{\rm FRB}}
\mathcal{L}_i({\Theta}),
\label{eq:mcmc_posterior}
\end{equation}
where $\mathcal{D}$ denotes the observed
FRB sample and
$\mathcal{L}_i({\Theta})$
is the host-marginalized likelihood
defined in Equation~\ref{eq:sz_marginal}.

\begin{figure*}[htbp]
    \centering
    \includegraphics[width=0.85\textwidth]{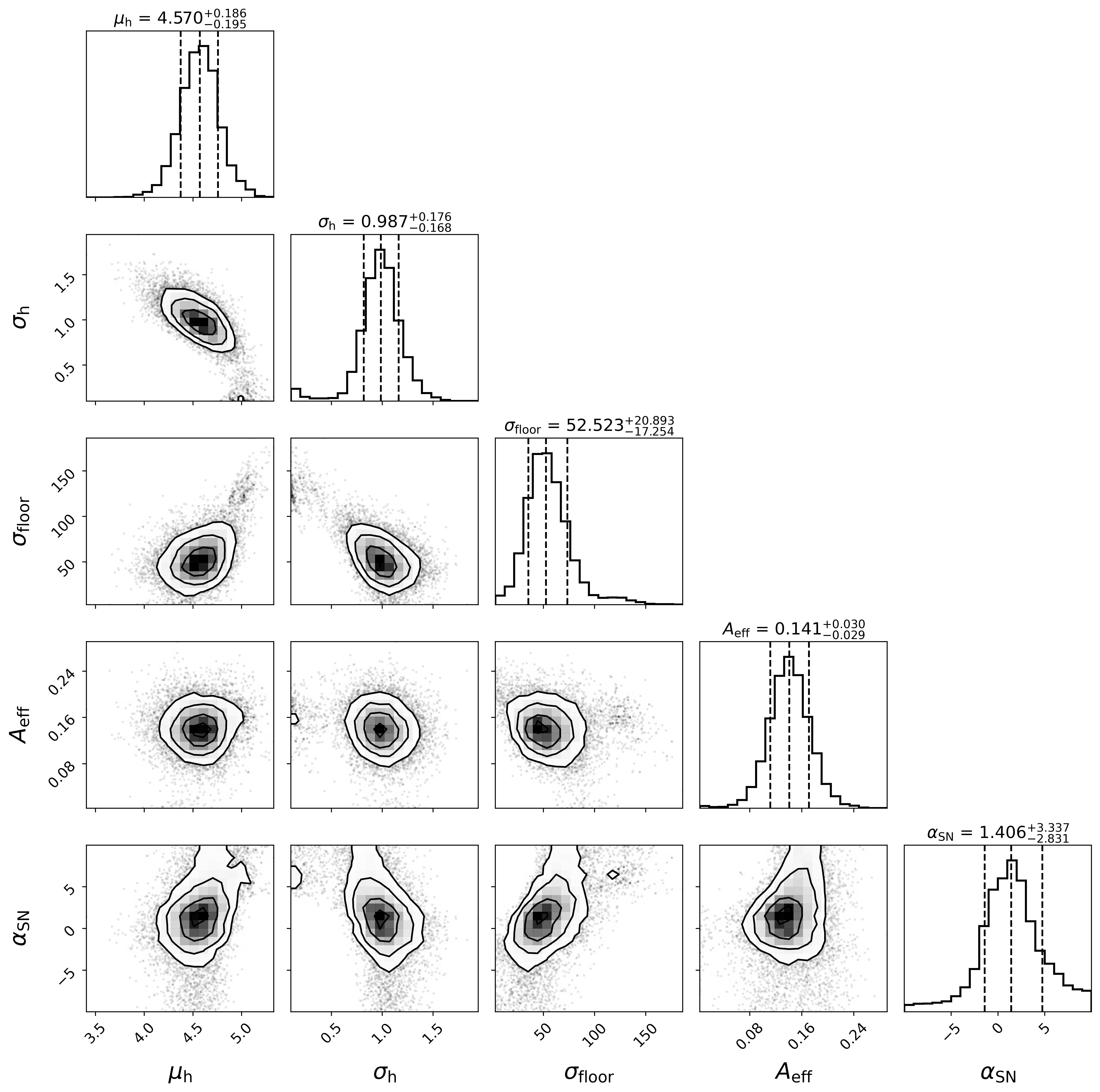}
    \caption{
    Marginalized posterior distributions
    and pairwise correlations of the five
    hyperparameters
    $(\mu_h,\sigma_h,\sigma_{\rm floor},
    A_{\rm eff},\alpha_{\rm SN})$
    obtained from the MCMC analysis of
    92 localized FRBs.
    The diagonal panels show the
    one-dimensional marginalized posterior
    distributions, with dashed vertical
    lines indicating the 16th, 50th,
    and 84th percentiles.
    The off-diagonal panels show the
    joint posterior distributions and
    their correlations.
    The titles above the diagonal panels
    report the posterior medians and
    corresponding 68\% credible intervals.
    }
    \label{fig:mcmc_corner}
\end{figure*}

Figure~\ref{fig:mcmc_corner} presents
the marginalized posterior distributions
and pairwise correlations of the five
parameters.
We found that the posterior medians are generally consistent with the maximum likelihood estimates obtained by the EM algorithm. Among the five parameters, $\alpha_{\rm SN}$ has the highest relative uncertainty. Its $68\%$ credible interval includes zero, indicating that the data do not provide strong evidence for nonzero residual skewness.
After marginalizing over the model parameters, the source-level probabilities $P(\Delta\mathrm{DM}_{\rm cos}>0)$ remain broadly consistent with the EM estimates, with a median absolute difference of 0.009. Nevertheless, several sources initially close to the adopted threshold of $P=0.7$ cross this threshold after marginalization. This indicates that the threshold-based candidate list is sensitive to small changes in the estimated probabilities.

\section{Raw data and calculation results}
\label{data}

Table~\ref{tab:frb} presents the information about 92 localized FRBs, including redshift, $\mathrm{DM}_{\rm obs}$, and the Milky Way interstellar-medium contribution $\mathrm{DM}_{\rm MW,ISM}$ estimated using NE2001. Following the common decomposition $\mathrm{DM}_{\rm MW}=\mathrm{DM}_{\rm MW,ISM}+\mathrm{DM}_{\rm MW,halo}$, we adopt a fixed $\mathrm{DM}_{\rm MW,halo}=30~{\rm pc~cm^{-3}}$ in this work. Table~\ref{tab:frb_dm} lists the inferred results for each FRB. The column $\langle \mathrm{DM}_{\rm cos}\rangle$ is the mean cosmological DM contribution evaluated at the source redshift using Equation~(\ref{eq:DMcos}). The column $\mathrm{DM}_{\rm host}$ gives the conditional mean of the inferred host-DM distribution (i.e., the mean host contribution under the fitted host population). The columns $\Delta\mathrm{DM}_{\rm cos}^{p16}$, $\Delta\mathrm{DM}_{\rm cos}^{p50}$, and $\Delta\mathrm{DM}_{\rm cos}^{p84}$ are the 16th, 50th, and 84th percentiles of the inferred $\Delta \mathrm{DM}_{\rm cos}$ distribution. The column $\Delta \mathrm{DM}_{\rm cos}$ is a point estimate obtained by evaluating Equation~(\ref{eq:exc}) with the conditional mean host contribution $ \mathrm{DM}_{\rm host}$; because it is based on the mean rather than the median of the host-DM inference, it may differ from the median column $\Delta\mathrm{DM}_{\rm cos}^{p50}$. The final column reports the model-conditioned probability $P(\Delta \mathrm{DM}_{\rm cos}>0)$.

\startlongtable
\begin{deluxetable*}{lcccc|lcccc}
\tabletypesize{\scriptsize}
\tablecaption{Information about 92 localized FRBs\label{tab:frb}}
\tablewidth{0pt}
\tablehead{
\colhead{Name} &
\colhead{$z$} &
\colhead{$\mathrm{DM}_{\rm obs}$} &
\colhead{$\mathrm{DM}_{\rm MW,ISM}$} &
\colhead{Cite}
&
\colhead{Name} &
\colhead{$z$} &
\colhead{$\mathrm{DM}_{\rm obs}$} &
\colhead{$\mathrm{DM}_{\rm MW,ISM}$} &
\colhead{Cite}
}
\startdata
FRB121102A & 0.19273 & 557 & 188.4  & 1,2 &
FRB171020A & 0.00867 & 114.1 & 36.7  & 4 \\
FRB180301A & 0.3304 & 552 & 151.7  & 3 &
FRB180814A & 0.068 & 190.9 & 87.6  & 16 \\
FRB180916B & 0.0337 & 349.35 & 199  & 5 &
FRB180924B & 0.3212 & 361.42 & 40.5 & 6,8,10,11,25 \\
FRB181030A & 0.00385 & 103.4 & 41.1  & 17 &
FRB181112A & 0.4755 & 589.27 & 41.7 & 9,10,11,25 \\
FRB181220A & 0.02746 & 208.66 & 118.5  & 18 &
FRB181223C & 0.03024 & 111.61 & 19.9  & 18 \\
FRB190102C & 0.2912 & 364.5 & 57.4 &  10,11,25 &
FRB190110C & 0.12244 & 221.6 & 37.1  & 19 \\
FRB190303A & 0.064 & 223.2 & 29.8 & 16 &
FRB190418A & 0.07132 & 182.78 & 70.2  & 18 \\
FRB190425A & 0.03122 & 127.78 & 48.7  & 18 &
FRB190520B & 0.2418 & 1204.7 & 60.2  & 6,12 \\
FRB190523A & 0.66 & 760.8 & 37.2 & 20 &
FRB190608B & 0.11778 & 338.7 & 37.3  & 6,10,11,25 \\
FRB190611B & 0.3778 & 321.4 & 57.8  & 7,10,25 &
FRB190614D & 0.6 & 959.2 & 87.8  & 21 \\
FRB190711A & 0.522 & 593.1 & 56.5  & 7,10,25 &
FRB190714A & 0.2365 & 504.13 & 38.5  & 7,25 \\
FRB191001A & 0.234 & 506.92 & 44.2 & 7,25 &
FRB191106C & 0.10775 & 332.2 & 25 & 19 \\
FRB191228A & 0.2432 & 297.5 & 32.9  & 3,25 &
FRB200223B & 0.06024 & 201.8 & 45.6  & 19 \\
FRB200430A & 0.1608 & 380.1 & 27.2  & 7,25 &
FRB200906A & 0.3688 & 577.8 & 35.8  & 3,25 \\
FRB201123A & 0.0507 & 433.55 & 251.7 & 22 &
FRB201124A & 0.098 & 413.52 & 139.9  & 13 \\
FRB210117A & 0.2145 & 729.1 & 34.4 & 6,25 &
FRB210320C & 0.2797 & 384.8 & 39.3  & 6,25 \\
FRB210405I & 0.066 & 565.17 & 516.1  & 26 &
FRB210410D & 0.1415 & 571.2 & 56.2  & 6,14 \\
FRB210603A & 0.1772 & 500.15 & 39.5  & 23 &
FRB210807D & 0.1293 & 251.9 & 121.2  & 6,25 \\
FRB211127I & 0.0469 & 234.83 & 42.5 & 6,15,25 &
FRB211203C & 0.3439 & 636.2 & 63.7  & 6,25 \\
FRB211212A & 0.0707 & 206 & 38.8  & 6,25 &
FRB220105A & 0.2785 & 583 & 22  & 6,25 \\
FRB220204A & 0.4012 & 612.58 & 50.7 & 24,28,29 &
FRB220207C & 0.04304 & 262.38 & 76.1  & 24,29 \\
FRB220307B & 0.24812 & 499.27 & 128.2  & 24,29 &
FRB220310F & 0.47796 & 462.24 & 46.3  & 24,29 \\
FRB220319D & 0.01123 & 110.98 & 139.7& 24 &
FRB220418A & 0.622 & 623.25 & 36.7 &  24,29 \\
FRB220501C & 0.381 & 449.5 & 30.6 & 25,28 &
FRB220506D & 0.30039 & 396.97 & 84.6 & 24,28,29 \\
FRB220509G & 0.0894 & 269.53 & 55.6 &  18,24,29 &
FRB220529A & 0.1839 & 246 & 40  & 27 \\
FRB220610A & 1.016 & 1458.15 & 31.0  & 25 &
FRB220717A & 0.36295 & 637.34 & 118.3  & 32 \\
FRB220725A & 0.1926 & 290.4 & 30.7  & 25 &
FRB220726A & 0.3619 & 686.23 & 89.5  & 24,28,29 \\
FRB220825A & 0.2414 & 651.24 & 78.5 & 24,29 &
FRB220831A & 0.262 & 1146.25 & 126.8 & 29 \\
FRB220912A & 0.0771 & 219.46 & 125.2  & 31 &
FRB220914A & 0.1139 & 631.28 & 54.7  & 24,29 \\
FRB220918A & 0.491 & 656.8 & 153.1  & 25 &
FRB220920A & 0.15824 & 314.99 & 39.9  & 24,29 \\
FRB221012A & 0.28467 & 441.08 & 54.3  & 24,29 &
FRB221029A & 0.975 & 1391.75 & 43.8  & 24,28,29 \\
FRB221101B & 0.2395 & 491.55 & 131.2  & 24,28,29 &
FRB221106A & 0.2044 & 343.8 & 34.8  & 24,25 \\
FRB221113A & 0.2505 & 411.03 & 91.7  & 24,28,29 &
FRB221116A & 0.2764 & 643.45 & 132.3  & 28,29 \\
FRB221219A & 0.553 & 706.71 & 44.4  & 24,28,29 &
FRB230124A & 0.0939 & 590.574 & 38.6  & 24,28,29 \\
FRB230307A & 0.2706 & 608.85 & 37.6  & 24,28,29 &
FRB230501A & 0.3015 & 532.47 & 125.7 & 24,29 \\
FRB230521B & 1.354 & 1342.9 & 138.8 & 25,29 &
FRB230526A & 0.157 & 361.4 & 31.9  & 25 \\
FRB230626A & 0.327 & 452.72 & 39.3 & 24,28,29 &
FRB230628A & 0.127 & 344.95 & 39 & 24,28,29 \\
FRB230708A & 0.105 & 411.51 & 60.3  & 25 &
FRB230712A & 0.4525 & 587.57 & 39.2  & 24,28,29 \\
FRB230718A & 0.0357 & 477 & 420.6  & 25 &
FRB230814A & 0.553 & 696.4 & 104.8  & 29 \\
FRB230902A & 0.3619 & 440.1 & 34.1  & 25 &
FRB231120A & 0.0368 & 437.74 & 43.8  & 24,28,29 \\
FRB231123B & 0.2621 & 396.86 & 40.3  & 24,28,29 &
FRB231220A & 0.3355 & 491.2 & 49.9  & 29 \\
FRB231226A & 0.1569 & 329.9 & 38.1  & 25 &
FRB240114A & 0.13 & 527.65 & 49.7  & 30 \\
FRB240119A & 0.376 & 483.1 & 38  & 29 &
FRB240123A & 0.968 & 1462.0 & 90.2  & 29 \\
FRB240201A & 0.04273 & 374.5 & 38.6  & 25 &
FRB240210A & 0.02369 & 283.73 & 28.7  & 25 \\
FRB240213A & 0.1185 & 357.4 & 40  & 29 &
FRB240215A & 0.21 & 549.5 & 47.9  & 29 \\
FRB240229A & 0.287 & 491.15 & 38  & 29 &
FRB240310A & 0.127 & 601.8 & 30.1  & 25 \\
\enddata
\tablecomments{The unit of each DM is ${\rm pc\ cm^{-3}}$. The listed $\mathrm{DM}_{\rm MW,ISM}$ values are estimated using NE2001. The references in the Table~\ref{tab:frb} are as follows: 1, \cite{chatterjee_direct_2017}, 2, \cite{tendulkar_host_2017}, 3, \cite{bhandari_characterizing_2022}, 4, \cite{mahony_search_2018}, 5, \cite{marcote_repeating_2020}, 6, \cite{gordon_demographics_2023}, 7, \cite{heintz_host_2020}, 8, \cite{doi:10.1126/science.aaw5903}, 9, \cite{doi:10.1126/science.aay0073}, 10, \cite{macquart_census_2020}, 11, \cite{bhandari_host_2020}, 12, \cite{niu_repeating_2022}, 13, \cite{fong_chronicling_2021}, 14, \cite{caleb_subarcsec_2023}, 15, \cite{glowacki_wallaby_2023}, 16, \cite{michilli_subarcminute_2023}, 17, \cite{bhardwaj_local_2021}, 18, \cite{bhardwaj_host_2024}, 19, \cite{ibik_proposed_2024}, 20, \cite{ravi_fast_2019}, 21, \cite{law_distant_2020}, 22, \cite{rajwade_first_2022}, 23, \cite{Cassanelli_2024}, 24, \cite{sherman_deep_2024}, 25, \cite{Shannon_2025}, 26, \cite{driessen_frb_2023}, 27, \cite{Gao_2025}, 28, \cite{sharma_preferential_2024}, 29, \cite{connor_gas-rich_2025}, 30, \cite{tian_detection_2024}, 31, \cite{ravi_deep_2023}, 32, \cite{rajwade_study_2024}.}
\end{deluxetable*}

\startlongtable
\begin{deluxetable*}{lrrrrrrr}
\tabletypesize{\scriptsize}
\tablecaption{Calculation results of 92 localized FRBs\label{tab:frb_dm}}
\tablewidth{0pt}
\tablehead{
\colhead{Name} &
\colhead{$\langle \mathrm{DM}_{\rm cos}\rangle$} &
\colhead{$\mathrm{DM}_{\rm host}$} &
\colhead{$\Delta \mathrm{DM}_{\rm cos}$} &
\colhead{$\Delta\mathrm{DM}_{\rm cos}^{p16}$} &
\colhead{$\Delta\mathrm{DM}_{\rm cos}^{p50}$} &
\colhead{$\Delta\mathrm{DM}_{\rm cos}^{p84}$} &
\colhead{$P(\Delta \mathrm{DM}_{\rm cos}>0)$}
}
\startdata
FRB121102A & 190.4871 & 134.5059 & 35.3414 & -31.3558 & 42.2968 & 101.2993 & 0.7159 \\
FRB171020A & 8.2149 & 59.0649 & -19.3721 & -49.9884 & -15.7745 & 11.0266 & 0.2956 \\
FRB180301A & 334.2197 & 87.0392 & -29.3430 & -71.1491 & -20.3837 & 10.9603 & 0.2845 \\
FRB180814A & 65.4069 & 53.2500 & -41.9665 & -69.5633 & -37.4779 & -14.8332 & 0.0115 \\
FRB180916B & 32.1364 & 86.6243 & 4.4124 & -38.8066 & 7.7384 & 47.8496 & 0.5664 \\
FRB180924B & 324.4744 & 64.1288 & -82.0928 & -111.8542 & -75.2740 & -53.0510 & 0 \\
FRB181030A & 3.6423 & 54.0019 & -25.1411 & -53.2936 & -21.5905 & 2.7594 & 0.1956 \\
FRB181112A & 489.7448 & 94.1897 & -36.0106 & -78.5144 & -25.5206 & 4.8013 & 0.2176 \\
FRB181220A & 26.1433 & 58.8977 & -23.3070 & -53.9229 & -19.3341 & 6.9828 & 0.2465 \\
FRB181223C & 28.8110 & 58.7307 & -24.1078 & -54.6438 & -20.0833 & 6.0889 & 0.2356 \\
FRB190102C & 292.8195 & 66.6673 & -67.3516 & -98.9748 & -60.3920 & -36.5539 & 0 \\
FRB190110C & 119.2695 & 69.1010 & -26.3327 & -61.7447 & -20.6598 & 8.3520 & 0.2573 \\
FRB190303A & 61.4992 & 97.1816 & 10.5647 & -37.7588 & 14.3824 & 59.0690 & 0.6085 \\
FRB190418A & 68.6557 & 55.7507 & -38.1150 & -66.9897 & -33.5095 & -9.7270 & 0.0495 \\
FRB190425A & 29.7522 & 53.3995 & -32.4551 & -60.3227 & -28.4474 & -4.9548 & 0.0988 \\
FRB190520B & 241.1531 & 1042.4323 & 33.8943 & -58.7062 & 25.3471 & 128.0644 & 0.6091 \\
FRB190523A & 690.2670 & 97.4689 & -55.3832 & -95.3910 & -44.1538 & -16.8128 & 0.0017 \\
FRB190608B & 114.6117 & 137.7424 & 33.5598 & -31.3924 & 36.8525 & 99.5239 & 0.7003 \\
FRB190611B & 384.6794 & 45.3345 & -183.9829 & -203.3900 & -179.1907 & -164.7769 & 0 \\
FRB190614D & 624.9029 & 186.9800 & 99.6346 & 9.3905 & 121.4302 & 182.0600 & 0.8575 \\
FRB190711A & 540.1093 & 78.8178 & -85.2950 & -118.8545 & -76.5409 & -52.6680 & 0 \\
FRB190714A & 235.6473 & 174.7333 & 58.6699 & -24.2831 & 65.1618 & 141.6791 & 0.7626 \\
FRB191001A & 233.0530 & 174.4564 & 58.2923 & -24.4442 & 64.6656 & 141.1471 & 0.7616 \\
FRB191106C & 104.6153 & 150.0312 & 37.1469 & -30.5609 & 38.8808 & 106.6666 & 0.7053 \\
FRB191228A & 242.6087 & 64.9093 & -60.2202 & -91.7324 & -53.6380 & -29.5084 & 0 \\
FRB200223B & 57.8325 & 78.2095 & -5.3983 & -45.4332 & -0.9459 & 34.3383 & 0.4910 \\
FRB200430A & 157.9213 & 145.4341 & 39.6908 & -29.9756 & 44.4811 & 109.8210 & 0.7206 \\
FRB200906A & 375.0683 & 135.4072 & 38.0077 & -29.2977 & 50.8074 & 101.8881 & 0.7429 \\
FRB201123A & 48.5580 & 97.0439 & 10.9308 & -36.9116 & 14.2677 & 59.1434 & 0.6089 \\
FRB201124A & 94.9368 & 130.8443 & 29.5171 & -32.1613 & 32.4279 & 92.2557 & 0.6871 \\
FRB210117A & 212.8777 & 481.1230 & 55.6733 & -42.2298 & 46.0810 & 155.2792 & 0.6872 \\
FRB210320C & 280.7384 & 82.6343 & -29.8115 & -70.2055 & -21.6233 & 9.2566 & 0.2656 \\
FRB210405I & 63.4522 & 37.9673 & -79.9988 & -99.5122 & -76.2946 & -60.7665 & 0 \\
FRB210410D & 138.4070 & 339.0840 & 49.5418 & -36.9310 & 41.6692 & 138.2807 & 0.6902 \\
FRB210603A & 174.6056 & 229.9747 & 60.6840 & -28.6657 & 58.5818 & 154.1536 & 0.7397 \\
FRB210807D & 126.1416 & 49.2296 & -69.0346 & -93.8533 & -64.2378 & -44.6781 & 0 \\
FRB211127I & 44.8755 & 105.9449 & 16.2558 & -34.6307 & 18.8430 & 67.9293 & 0.6347 \\
FRB211203C & 348.5497 & 170.9008 & 66.7825 & -17.8037 & 79.0467 & 148.1931 & 0.7918 \\
FRB211212A & 68.0487 & 79.5713 & -5.1657 & -45.9179 & -0.4720 & 35.2196 & 0.4956 \\
FRB220105A & 279.4795 & 221.5484 & 78.2327 & -20.4705 & 82.9298 & 178.7078 & 0.7833 \\
FRB220204A & 409.7280 & 129.0825 & 30.0331 & -33.3151 & 43.1537 & 89.9569 & 0.7230 \\
FRB220207C & 41.1417 & 104.1407 & 15.2948 & -34.7895 & 17.8553 & 66.1471 & 0.6299 \\
FRB220307B & 247.7315 & 105.7850 & 8.5833 & -44.7269 & 17.5387 & 60.1013 & 0.6186 \\
FRB220310F & 492.4024 & 59.5078 & -146.7259 & -171.5769 & -140.3106 & -122.2756 & 0 \\
FRB220319D & 10.6436 & 27.1350 & -96.1973 & -110.3484 & -93.5008 & -82.1899 & 0 \\
FRB220418A & 648.8622 & 71.5500 & -136.4244 & -164.6867 & -128.6291 & -108.6877 & 0 \\
FRB220501C & 388.0999 & 78.7598 & -56.2308 & -92.5676 & -47.7611 & -21.0490 & 0 \\
FRB220506D & 302.4957 & 66.1475 & -70.9931 & -102.1812 & -64.0461 & -40.6080 & 0 \\
FRB220509G & 86.4324 & 96.4960 & 8.9205 & -39.7037 & 13.6527 & 57.3564 & 0.6023 \\
FRB220529A & 181.4477 & 60.3759 & -56.4452 & -86.4679 & -50.5593 & -27.1141 & 0 \\
FRB220610A & 1076.0596 & 240.5352 & 201.7773 & 97.2926 & 237.6797 & 293.6647 & 0.9352 \\
FRB220717A & 368.8283 & 126.2186 & 27.6048 & -34.9078 & 39.8114 & 87.0274 & 0.7116 \\
FRB220725A & 190.3539 & 77.1663 & -25.3582 & -64.2238 & -18.4330 & 12.4632 & 0.2999 \\
FRB220726A & 367.7089 & 174.7774 & 70.6896 & -15.7769 & 83.6789 & 153.6064 & 0.7990 \\
FRB220825A & 240.7342 & 280.2348 & 76.2644 & -27.2162 & 72.8919 & 184.9746 & 0.7577 \\
FRB220831A & 262.2059 & 864.0063 & 42.6115 & -55.3239 & 33.3425 & 142.1500 & 0.6362 \\
FRB220912A & 74.3234 & 48.3446 & -54.9474 & -79.8850 & -50.5673 & -30.4374 & 0 \\
FRB220914A & 110.7400 & 445.1140 & 36.2404 & -42.0535 & 28.7356 & 115.7866 & 0.6466 \\
FRB220918A & 506.5128 & 76.9212 & -84.4031 & -117.6120 & -75.8996 & -52.1140 & 0 \\
FRB220920A & 155.3243 & 97.5010 & 5.5854 & -44.0635 & 12.3251 & 54.2507 & 0.5909 \\
FRB221012A & 285.9547 & 97.7727 & -5.2820 & -53.7917 & 3.9555 & 41.4291 & 0.5314 \\
FRB221029A & 1032.0302 & 220.8828 & 174.0804 & 78.5006 & 206.4359 & 259.0535 & 0.9269 \\
FRB221101B & 238.7628 & 104.3231 & 7.4257 & -45.2159 & 16.1566 & 58.3571 & 0.6113 \\
FRB221106A & 202.4717 & 94.4208 & -1.8683 & -49.7057 & 5.8787 & 44.6187 & 0.5464 \\
FRB221113A & 250.2072 & 81.9570 & -26.4196 & -66.9644 & -18.5629 & 12.8538 & 0.3036 \\
FRB221116A & 277.2774 & 178.1041 & 64.3343 & -21.4916 & 72.8054 & 149.1881 & 0.7768 \\
FRB221219A & 573.7725 & 109.6755 & -12.0862 & -61.0232 & 0.5748 & 34.4626 & 0.5048 \\
FRB230124A & 90.8785 & 435.5310 & 32.9503 & -41.4839 & 25.8224 & 108.5878 & 0.6387 \\
FRB230307A & 271.2008 & 241.3881 & 80.0736 & -22.0323 & 82.1647 & 185.5165 & 0.7783 \\
FRB230501A & 303.6657 & 99.9709 & -3.7067 & -53.1144 & 5.9005 & 43.7775 & 0.5456 \\
FRB230521B & 1432.5770 & 79.4309 & -292.2199 & -314.3181 & -284.9117 & -270.0502 & 0 \\
FRB230526A & 154.0687 & 131.0849 & 32.1341 & -32.2385 & 37.7304 & 96.3839 & 0.7028 \\
FRB230626A & 330.6162 & 93.2480 & -17.4630 & -62.7170 & -8.0289 & 26.0552 & 0.4266 \\
FRB230628A & 123.8355 & 134.5863 & 32.6965 & -31.7303 & 36.6777 & 97.7870 & 0.6998 \\
FRB230708A & 101.8816 & 192.9630 & 44.7013 & -30.5359 & 42.5094 & 123.2586 & 0.7098 \\
FRB230712A & 464.9065 & 101.9277 & -16.7135 & -63.9309 & -5.4750 & 28.3866 & 0.4516 \\
FRB230718A & 34.0612 & 44.5504 & -50.6760 & -73.9594 & -46.8343 & -27.7393 & 0 \\
FRB230814A & 573.7725 & 86.6710 & -67.9813 & -104.8695 & -58.2314 & -32.2841 & 0 \\
FRB230902A & 367.7089 & 79.7223 & -50.2464 & -87.4757 & -41.7607 & -14.2268 & 0.0173 \\
FRB231120A & 35.1207 & 311.1125 & 28.7464 & -36.3189 & 22.5764 & 95.0161 & 0.6385 \\
FRB231123B & 262.3104 & 93.2300 & -9.6223 & -56.0570 & -0.9607 & 35.2123 & 0.4919 \\
FRB231220A & 339.6292 & 101.6776 & -4.4636 & -54.1528 & 5.7034 & 43.1571 & 0.5441 \\
FRB231226A & 153.9673 & 107.3218 & 15.0660 & -39.2594 & 21.6451 & 68.5394 & 0.6417 \\
FRB240114A & 126.8438 & 307.4932 & 48.9883 & -35.5208 & 41.6973 & 136.0771 & 0.6936 \\
FRB240119A & 382.7561 & 89.0083 & -32.3424 & -74.2663 & -22.8882 & 8.0133 & 0.2530 \\
FRB240123A & 1024.4988 & 240.7955 & 194.9457 & 88.1548 & 230.2519 & 288.7016 & 0.9302 \\
FRB240201A & 40.8412 & 241.5531 & 33.4041 & -33.6498 & 27.6702 & 102.3495 & 0.6627 \\
FRB240210A & 22.5279 & 176.3240 & 30.2579 & -31.2322 & 26.6281 & 94.1723 & 0.6667 \\
FRB240213A & 115.3308 & 149.8621 & 38.0843 & -30.3944 & 40.3709 & 108.1747 & 0.7093 \\
FRB240215A & 208.2376 & 236.8185 & 67.6447 & -26.8288 & 66.5850 & 166.2807 & 0.7535 \\
FRB240229A & 288.4037 & 130.4381 & 33.3958 & -32.1804 & 43.8485 & 96.5370 & 0.7218 \\
FRB240310A & 123.8355 & 425.7676 & 40.0760 & -41.2605 & 32.2093 & 122.8158 & 0.6580 \\
\enddata
\tablecomments{The unit of each DM quantity is ${\rm pc\ cm^{-3}}$. Column definitions are given in the text above the table.}
\end{deluxetable*}

\bibliography{arxiv}{}
\bibliographystyle{aasjournal}

\end{document}